\documentclass{ptapap}

\def \ptaJournalDate{to be published}
\def \ptaJournalVolume{}
\def \ptaArticleURL{}
\def \ptaJournalName{PTA Proceedings}
\def \ArticleFirstPage{1}
\author{Ewa L. {\L}okas}[CAMK]
\author{Marcin Semczuk}[OAUJ]
\affil[CAMK]{Nicolaus Copernicus Astronomical Center, Bartycka 18, 00--716 Warszawa, Poland}
\affil[OAUJ]{Astronomical Observatory of the Jagiellonian University, Orla 171, 30--244 Krak{\'o}w, Poland}

\title{Tidal evolution of disky dwarf galaxies: prograde versus retrograde orbits}

\begin{document}

\maketitle

\begin{abstract}
The formation of dwarf spheroidal galaxies in the Local Group from disky progenitors via tidal interaction
with a bigger host is one of the most promising scenarios of their origin. Using $N$-body simulations we
study the process by following the evolution of a disky dwarf orbiting a Milky Way-like host. We focus on the
effect of the orientation of the dwarf galaxy disk's angular momentum with respect to the orbital one. We
find a strong dependence of the efficiency of the transformation from a disk to a spheroid on the disk
orientation. The effect is strongest for the exactly prograde and weakest for the exactly retrograde orbit.
In the prograde case the stellar component forms a strong bar and remains prolate until the end of the
evolution, while its rotation is very quickly replaced by random motions of the stars. In the retrograde case
the dwarf remains oblate, does not form a bar and loses rotation very slowly. Our results suggest that
resonant effects are the most important mechanism underlying the evolution while tidal shocking plays only a
minor role.
\end{abstract}

\section{Introduction}

The formation of dwarf spheroidal (dSph) galaxies in the Local Group via the tidal interaction of their disky
progenitors with more massive hosts like the Milky Way is one of the most promising scenarios for their origin
\citep{mayer01}. It explains the morphology-density relation observed among the dwarfs of the Local Group and
accounts for the non-sphericity of the dSph objects. The efficiency of the mechanism and its observational
predictions have been investigated in detail by \citet{kliment09}, \citet{kazan11} and \citet{lokas11,
lokas12}. These studies explored the dependence of the process on a large number of orbital and structural
parameters of the dwarf.

One such parameter expected to have a strong impact on the evolution is the inclination between the angular
momentum of the dwarf galaxy disk and its orbital angular momentum. However, in the studies mentioned above
only a narrow range of inclinations were studied in detail, namely those with values $i = 0^\circ$,
$45^\circ$ and $90^\circ$. Because of this range, and the way the properties of the dwarf were measured, no
clear evidence for the dependence on this parameter was found.

In this paper we report preliminary results of a new study aimed at clarifying this issue. For this
purpose we performed three simulations of tidal evolution of a dwarf galaxy with disk inclinations $i =
0^\circ$ (exactly prograde), $90^\circ$ (intermediate) and $180^\circ$ (exactly retrograde). We also measured
the properties of the dwarf galaxy in a different way that enables clear comparisons. We find a strong
dependence of the evolution on the initial inclination of the disk.

\section{The simulations}

We simulated the interaction between the dwarf galaxy and the host using two live, two-component galaxy
models. The simulation setup was similar to that in \citet{lokas14}. Our dwarf had a standard NFW
\citep{NFW} dark matter halo of mass $M_{\rm h}=10^9$ M$_{\odot}$ and a concentration parameter $c=20$. Its
disk had a mass $M_{\rm d}=2 \times 10^7$ M$_{\odot}$, an exponential scale-length $R_{\rm d}=0.41$ kpc and a
thickness $z_{\rm d}=0.2\ R_{\rm d}$. The host galaxy had the properties similar to the Milky Way model MWb
of \citet{widrow}. It had a dark matter halo of mass $M_{\rm H}=7.7\times 10^{11}$ M$_{\odot}$ and
concentration $c=27$ while its disk had a mass $M_{\rm D} = 3.4 \times 10^{10}$ M$_{\odot}$, a length-scale
$R_{\rm D}=2.82$ kpc and a thickness $z_{\rm D}=0.44$ kpc. We neglect other structural parameters of the
Milky Way, like the bulge, the thin/thick disk or a bar.

The numerical realizations of the two galaxies were generated by the procedures described in \citet{widrow}
and \citet{widrowBlueprints}. Each object was modeled with $2\times 10^5$ particles per component ($8\times
10^5$ particles in total). The evolution of the system
was followed with the GADGET-2 $N$-body code (\citealt{springelNA}; \citealt{springelCode}) adopting
gravitational softening scales of $\epsilon_{\rm d}=0.02$ kpc and $\epsilon_{\rm h}=0.06$ kpc for the dwarf's
disk and halo and $\epsilon_{\rm D}=0.05$ kpc and $\epsilon_{\rm H}=2$ kpc for the Milky Way, respectively.
The simulation lasted for 10 Gyr and outputs were saved every $0.05$ Gyr.

The initial configuration for all three simulations was such that the dwarf's orbit and the Milky Way disk
were in the $XY$ plane of the simulation box. The dwarf galaxy was initially placed at the apocenter $(X, Y,
Z) = (-120, 0, 0)$ kpc of the orbit (which had a pericenter of 25 kpc) with a systemic velocity towards the
negative $Y$ axis, so that the orbital angular momentum is pointing towards the positive $Z$ of the
simulation box. The orientation of the (unit) angular momentum of the dwarf was $(L_X, L_Y, L_Z) = (0, 0, 1)$
for the exactly prograde case ($i = 0^\circ$), $(0, 0, -1)$ for the exactly retrograde case ($i = 180^\circ$)
and $(0, -1, 0)$ for the intermediate case ($i = 90^\circ$).

\begin{figure}[t]
\begin{center}
\includegraphics[width=0.6 \textwidth]{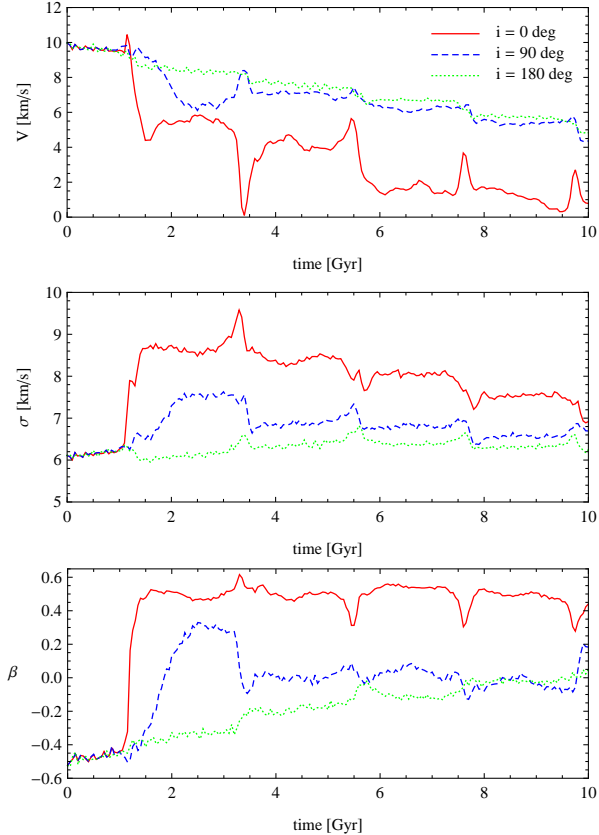}
\caption{Evolution of the kinematic properties of the dwarf galaxy in time. The upper panel shows the
mean rotation velocity $V$, the middle panel the average 1D velocity dispersion
$\sigma$ and the lower panel the anisotropy parameter $\beta$ of the stars within 0.5 kpc.
In each panel the solid, dashed and dotted line corresponds to the results for the
prograde ($i = 0^\circ$), intermediate ($i = 90^\circ$) and retrograde ($i = 180^\circ$) inclination of the
disk.}
\label{comparisonikinbeta}
\end{center}
\end{figure}

\section{Evolution of the kinematics and shape}

To quantify the kinematic evolution of the dwarf galaxy in each of the three simulations, for each output we
selected stars within the radius of 0.5 kpc and determined the principal axes of the stellar component from
the inertia tensor. Next, we rotated the stars so that the $x$ coordinate was along the longest, the $y$
along the intermediate, and $z$ along the shortest principal axis. A standard spherical coordinate system was
then introduced and we measured the kinematics using these coordinates.

The dominant component of the
streaming motion of the stars is always around the shortest axis. We measured the mean rotation around this
axis for stars within 0.5 kpc for all outputs of the three simulations. The results are shown in the upper
panel of Figure~\ref{comparisonikinbeta}. The middle panel of the Figure shows the measurements of the 1D velocity
dispersion which is the average of dispersions along the three spherical coordinates. In the lower panel we
plot the anisotropy parameter $\beta$ of the stars.

In each panel the solid, dashed and dotted line shows the results for the prograde ($i = 0^\circ$),
intermediate ($i = 90^\circ$) and retrograde ($i = 180^\circ$) inclination of the disk. Abrupt changes of
velocity and dispersion take place at pericenter passages that occur at $t = 1.2$, 3.3, 5.5, 7.6, 9.7 Gyr
from the start of the simulation. We find a systematic trend in the behaviour of both the rotation and
dispersion values among the three simulations: in the prograde case the rotation velocity decreases and the
dispersion grows most significantly in time and in the retrograde case the changes are the smallest.
Therefore the transition from the ordered to the random motion is strongest in the prograde case. In addition, in
this case the random motions are dominated by the radial velocity dispersion, as demonstrated by the highest
values of anisotropy of stellar orbits.

\begin{figure}[t]
\begin{center}
\includegraphics[width=0.6 \textwidth]{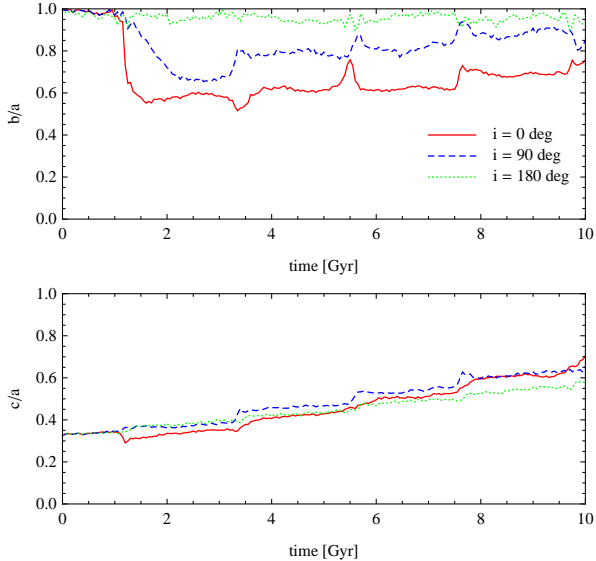}
\caption{Evolution of the shape of the dwarf galaxy in time in terms of the axis ratios of the stellar
component. The upper panel shows the intermediate to longest ($b/a$) and the lower one the
shortest to longest ($c/a$) axis ratio of stars within radius 0.5 kpc. In each panel the solid, dashed and
dotted line corresponds to the results for the prograde ($i = 0^\circ$), intermediate ($i = 90^\circ$) and
retrograde ($i = 180^\circ$) inclination of the disk.}
\label{comparisonishape}
\end{center}
\end{figure}

\begin{figure}[t]
\begin{center}
\includegraphics[width=0.6 \textwidth]{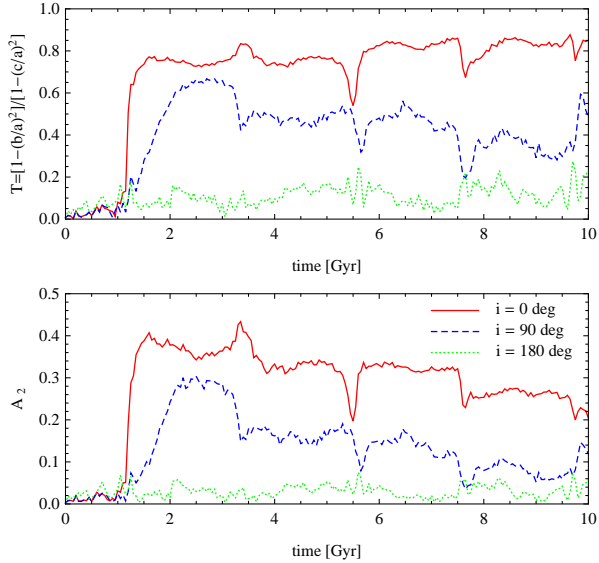}
\caption{Evolution of the shape of the dwarf galaxy in time in terms of the triaxiality parameter $T$ (upper
panel) and the bar mode $A_2$ (lower panel) of the stellar component.
In each panel the solid, dashed and dotted line corresponds to the results for the prograde ($i = 0^\circ$),
intermediate ($i = 90^\circ$) and retrograde ($i = 180^\circ$) inclination of the disk.}
\label{comparisonitria2}
\end{center}
\end{figure}

We also determined the shape of the stellar component from the same subsamples of stars. The results in terms
of the intermediate to longest $b/a$ and the shortest to longest $c/a$ axis ratio are shown in the upper and
lower panel of Figure~\ref{comparisonishape}, respectively. To further illustrate the evolution of the shape,
in Figure~\ref{comparisonitria2} we plot the triaxiality parameter
$T=[1-(b/a)^2]/[1-(c/a)^2]$ (upper panel) and the bar mode $A_2$ of the Fourier decomposition of the stellar
component projected along the shortest axis (lower panel). As before, the solid, dashed and dotted line shows
the results for the prograde, intermediate and retrograde orientation of the disk. Also for these quantities
most significant changes occur at pericenter passages.

While the evolution of $c/a$ in time is similar
in all cases, the variation of $b/a$ in time is very different for the prograde, intermediate and retrograde
orientation of the disk. In the prograde case the shape is always prolate  ($T > 2/3$), in the intermediate
case it is triaxial ($1/3 < T < 2/3$) and in the retrograde case it remains oblate ($T < 1/3$), like the
initial disk. In the retrograde case the dwarf does not undergo any significant morphological transformation,
it remains disky, only the disk thickens with time due to tidal shocks at pericenter passages. In the two
other cases, the dwarf forms a bar at the first pericenter passage, but the bar is much stronger in the
exactly prograde case, as confirmed by the highest $A_2$ values for this case at all times.

\section{Discussion}

We presented the results of three simulations of tidal evolution of initially disky dwarf galaxies orbiting
the Milky Way-like host. In all cases the dwarf was placed on a typical, eccentric orbit and its evolution
was followed for 10 Gyr. The three configurations differed only by the initial inclination of the dwarf
galaxy disk's angular momentum with respect to the orbital angular momentum. We find very significant
differences between the properties of the evolving dwarf for the three inclinations and they seem to scale
monotonically with the inclination angle.

The evolution of the dwarf is strongest in the case of prograde inclination of the disk. In this
configuration a strong bar forms at the first pericenter passage and although the shape of the stellar
component becomes more spherical in time, it remains prolate until the end of the simulation. The
morphological transformation is accompanied by strong modification of the dwarf's kinematics: at the first
pericenter the rotation is significantly reduced and replaced by random motions of the stars, mostly in the
radial direction as is characteristic of the bar. The rotation continues to decrease in time so that almost
no streaming motion remains at the end of the evolution. On the other hand, the decrease of the velocity
dispersion at later times reflects the mass loss due to tidal stripping.

In the intermediate case, that of perpendicular orientation of the disk with respect to the orbit, the
changes are qualitatively similar, but less pronounced. A bar also forms at the first pericenter passage, but
it is more triaxial than prolate. The decrease of rotation velocity and the increase of velocity dispersion
are also weaker effects. In the exactly retrograde case no significant evolution is seen: the dwarf's stellar
component does not form a bar and remains disky. The disk thickens in time and rotation is slightly
diminished.

The results presented here suggest that the most important mechanism underlying the evolution of disky dwarfs
is of resonant nature since the effect is strongest when the dwarf's angular momentum is aligned with the
orbital one. We may refer to it as `resonant stirring' in analogy to the `resonant stripping' mechanism found
to increase the mass loss in similar configurations \citep{donghia09, donghia10}. We plan to investigate the
details of the process in future work.

\section*{Acknowledgements}
This work was supported in part by PL-Grid Infrastructure, the Polish National Science Centre under
grant 2013/10/A/ST9/00023, by US National Science Foundation Grant No. PHYS-1066293 and the hospitality of
the Aspen Center for Physics. MS acknowledges the summer student program of the Copernicus Center. We thank L.
Widrow for providing procedures to generate $N$-body models of galaxies for initial conditions.

\bibliographystyle{ptapap}
\bibliography{pta-lokas-astroph}

\end{document}